\documentclass[12pt,preprint,aps,english,superscriptaddress]{revtex4-1}
\usepackage[T1]{fontenc}
\usepackage[latin9]{inputenc}
\usepackage{amsmath}
\usepackage{amssymb}
\usepackage{cancel}
\usepackage{graphicx}
\usepackage{esint}
\usepackage{color}
\makeatletter

\@ifundefined{textcolor}{}{%
 \definecolor{BLACK}{gray}{0}
 \definecolor{WHITE}{gray}{1}
 \definecolor{RED}{rgb}{1,0,0}
 \definecolor{GREEN}{rgb}{0,1,0}
 \definecolor{BLUE}{rgb}{0,0,1}
 \definecolor{CYAN}{cmyk}{1,0,0,0}
 \definecolor{MAGENTA}{cmyk}{0,1,0,0}
 \definecolor{YELLOW}{cmyk}{0,0,1,0}
}

\usepackage{dcolumn}
\usepackage{bm}
\usepackage{enumerate}

\@ifundefined{showcaptionsetup}{}{%
\PassOptionsToPackage{caption=false}{subfig}}

\usepackage{babel}
\addto\captionsenglish{}
\renewcommand{\thefigure}{\@arabic\c@figure}

\makeatother

\begin{document}

\title{Transient superconductivity from electronic squeezing of optically pumped phonons}

\author{Dante M. Kennes}

\thanks{These two authors contributed equally}

\affiliation{Department of Physics, Columbia University, New York, NY, 10027,
USA}

\author{Eli Y. Wilner}

\thanks{These two authors contributed equally}

\affiliation{Department of Physics, Columbia University, New York, NY, 10027,
USA}

\author{David R. Reichman}

\affiliation{Department of Chemistry, Columbia University, New York, NY, 10027,
USA}

\author{Andrew J. Millis}

\affiliation{Department of Physics, Columbia University, New York, NY, 10027,
USA}

\begin{abstract}
{\bf Advances in light sources and time resolved spectroscopy have made it possible to excite specific atomic vibrations in solids and to observe the resulting changes in electronic properties but the mechanism by  which phonon excitation causes qualitative changes in electronic properties, has remained unclear.  Here we show that the dominant symmetry-allowed coupling  between electron density and dipole active modes  implies an electron density-dependent squeezing of the phonon state which  provides an attractive contribution to the electron-electron interaction, independent of the sign of the bare electron-phonon coupling and with a magnitude proportional to the degree of laser-induced phonon excitation. Reasonable  excitation amplitudes lead to non-negligible attractive interactions that may cause significant transient changes in electronic properties including  superconductivity. The mechanism is generically applicable to a wide range of systems, offering a promising route to manipulating and controlling electronic phase behavior in novel materials.
}
\end{abstract}
\maketitle

Strong mode-specific excitation of specific atomic vibrations (phonon modes) in solids \cite{wall2009ultrafast} has been shown \cite{merlin1997generating,garrett1997vacuum,rini2007control,forst2011nonlinear,he2016coherent}  to  drive  drastic  changes in collective electronic properties. Of the many effects observed, perhaps the most dramatic is the superconducting-like behavior observed at temperatures far above the equilibrium transition temperature in strongly irradiated $\mbox{Y}\mbox{Ba}_{2}\mbox{Cu}_{3}\mbox{O}_{6+x}$ \cite{mankowsky2015coherent} and $\mbox{K}_{3}\mbox{C}_{60}$ \cite{mitrano2016possible}.  Because  optically addressable phonon modes are dipole active (odd parity) zone center vibrations which typically do not couple linearly  to local electronic quantities such as the density or orbital occupancy, the mechanism by which phonon excitation changes electronic properties has not been clear, although many interesting proposals have been made  \cite{mankowsky2014nonlinear,cavalleri2014theory,singla2015thz,knap2015dynamical,komnik2016bcs,kim2016enhancing}.  Here we present a new and  general mechanism that naturally explains how optical phonon excitation can lead to changes in electronic properties. The key idea is that the dominant symmetry-allowed coupling  between electron density and dipole active modes  is quadratic, implying an electron density-dependent squeezing of the phonon state.  We show that this electron density dependence of the phonon squeezing  provides a sizable attractive contribution to the electron-electron interaction. The attractive contribution is independent of the sign of the bare electron-phonon coupling,  has a magnitude proportional to the degree of laser-induced phonon excitation, and is large enough that reasonable values of the excitation amplitude lead to non-negligible attractive interactions. The mechanism is generically applicable to a wide range of systems, offering a promising route to manipulating and controlling electronic phase behavior.

To demonstrate the mechanism we consider a minimal tight-binding model of electrons that can hop between sites of a lattice, are subject to electronic interactions, and are coupled quadratically to an optical phonon that is excited by an external field. The  dimensionality of the lattice   and the precise nature of the bare electronic  interactions will not be qualitatively important in what follows, but we will invoke realistic and explicit coupling parameters and dimensionality when making connection to recent experiments.  The optical mode we study can be considered as a proxy for the $T_{1u}$ phonons which  are the dominant optically excited modes in the $\mbox{K}_{3}\mbox{C}_{60}$ experiments of Mitrano {\em et al.} \cite{mitrano2016possible} and are known to couple quadratically to electrons.   The Hamiltonian may be written as  (circumflexes denote quantum operators in cases where the symbol could be confused with its classical meaning) 
\begin{equation}
H =  -\sum_{ij\sigma}J_{ij}c_{i\sigma}^{\dagger}c_{j\sigma}+U_{elec}\sum_{i}\hat{n}_{i\uparrow}\hat{n}_{i\downarrow}+\sum_{i}\left(\frac{K}{2}\hat{x}_{i}^{2}+\frac{1}{2M}\hat{p}_{i}^{2}\right)+ gK\sum_{i}\hat{n}_{i}\hat{x}_{i}^{2}.
\label{H}
\end{equation}
Here $g$ is a dimensionless measure of the quadratic electron-phonon coupling, $\hat{n}_i$ denotes the electron density operator on site $i$, namely $\hat{n}_i=\sum_{\sigma}c_{i\sigma}^{\dagger}c_{i\sigma}$, and $\hat{x}_i$ is the phonon displacement operator on site $i$.   The usual linear electron-phonon coupling term can be trivially incorporated but is suppressed here for clarity: we assume that it acts primarily to modify the strength of the hopping and electron-electron interaction terms in equilibrium, which may then be taken as effective parameters.   

The quadratic electron-phonon coupling implies a density-dependent change of the oscillator stiffness $K\rightarrow K(1+2gn)$ ($g>-\frac{1}{4}$ is required by stability) implying that the phonon frequency $\sqrt{K/M}$ becomes $\omega(n)=\omega_{0}\sqrt{1+2gn}$.  To see that the electron density dependence of the phonon frequency induces an attractive interaction we note that  the effective electron-electron interaction   $U^\star$ is the energy of a site with two electrons  plus the energy of an empty  site, minus twice the energy of the singly occupied site: $U^\star\equiv E(n=2)+E(n=0)-2E(n=1)$. Thus the interaction energy associated to a site with $n_{\rm B}$ excited phonons  is
\begin{eqnarray}
U^\star=  U_{elec}-\left(n_{\rm B}+\frac{1}{2}\right)\omega_{0}\left(2\sqrt{1+2g}-1-\sqrt{1+4g}\right)\label{Ueff}.
\end{eqnarray}
The downward concavity of the square root means that $\left(2\sqrt{1+2g}-1-\sqrt{1+4g}\right)$ is positive \emph{for either sign} of $g$. The magnitude of the induced interaction is  proportional to the phonon frequency, demonstrating the essentially quantum aspect of the effect and also to the number of excited phonons. the magnitude of the induced interaction is also proportional to the number of excited phonon quanta, which can be controlled by the pump fluence.  Thus, this simple mechanism offers a direct, robust and generic  means to controllably change the magnitude and even the sign of the electron-electron interaction by phonon excitation. 

To analyse this effect mathematically we note that the quadratic electron-phonon coupling  of equation~\eqref{H} gives rise to an electron density-dependent change of oscillator stiffness without a corresponding change of mass, in other words to a density-dependent \emph{squeezing} \cite{glauber1963coherent} of the oscillator states, in contrast to the usual linear (Holstein or Fr\"ohlich) electron-phonon coupling which leads to an electron density-dependent {\em shift} of the oscillator equilibrium position. Just as the linear coupling can be treated with a Lang-Firsov canonical transformation \cite{lang1963kinetic} that shifts the phonon coordinate, the quadratic coupling can be treated with a squeezing transformation $e^{\hat{S}}=e^{\frac{i}{2}\sum_{j}\zeta_{j}\left(\hat{x}_{j}\hat{p}_{j}+\hat{p}_{j}\hat{x}_{j}\right)}$ that rescales oscillator position $\hat{x}_{j}\rightarrow e^{\hat{S}}\hat{x}_{j}e^{-\hat{S}}=e^{\zeta_{j}}\hat{x}_{j}$ and momentum $\hat{p}_{j}\rightarrow e^{-\zeta_{j}}\hat{p}_{j}$ and transforms the Hamiltonian as  $H\rightarrow \tilde{H}  \equiv  e^{\hat{S}}He^{-\hat{S}}$. The squeezing parameter is $\zeta_{j}=-\frac{1}{4}\ln\left[1+2g\left(\hat{n}_{j\uparrow}+\hat{n}_{j\downarrow}\right)\right]$. 

The details of the transformation are given in the Supplementary Information. The transformation of the phonon and electron-electron interaction terms is straightforward. As in the Lang-Firsov case the transformation of the hopping $J_{ij}$ generates expressions involving inelastic (phonon pair creation/annihilation) processes. The nonlinear dependence of $\hat{S}$ on the phonon operators means that the standard 'disentangling' \cite{mahan2013many} formulas that simplify the Lang-Firsov case do not apply, while the  nonlinear dependence of $\hat{S}$ on electron density means that  the transformed hopping depends on the occupancies of the states between which the electron hops. However the physically relevant situation involves small values of the dimensionless coupling $g$, with interesting behavior occurring when the product of $g$  and the number of excited phonons $n_{\rm B}$ is large enough. In this situation we find by comparison to the exactly solvable two-site version of the model (See Supplementary Information) that inelastic effects lead primarily to a rapid decoherence of the initially prepared phonon state,  so that on experimentally relevant timescales one may consider the phonons to be characterized by a density matrix which is diagonal in the site occupation number with Poisson-distributed eigenvalues initially determined by the pump fluence and decaying slowly back to the equilibrium while the electronic physics is described by  a renormalized hopping $J_{ij}^\star=J_{ij}e^{-g^2(n_{\rm B}^2+2n_{\rm B}+1)/8}$ so the effective Hamiltonian becomes ($\beta^\dagger$ creates an eigenstate of the  squeezed phonon Hamiltonian)
\begin{eqnarray}\label{Happrox}
\tilde{H} & \rightarrow & H_{\rm eff}=-\sum_{\left\langle i,j\right\rangle \sigma}J^{\star}_{ij}c_{i\sigma}^{\dagger}c_{j\sigma}+\omega_{0}\sum_{i}
\left(\beta_{i}^{\dagger}\beta_{i}+\frac12\right)+\sum_{i}U^{\star}n_{i\uparrow}n_{i\downarrow}\nonumber \\
& + & \frac{g\omega_{0}}{2}\left(1-\frac{g}{2}\right)\sum_{i \sigma}\left(2\beta_{i}^{\dagger}\beta_{i}+1\right)
n_{i\sigma},
\end{eqnarray}
with  $U^\star$ given by equation~\eqref{Ueff}.

\begin{figure}[t]
\centering{}\includegraphics[width=0.6\columnwidth]{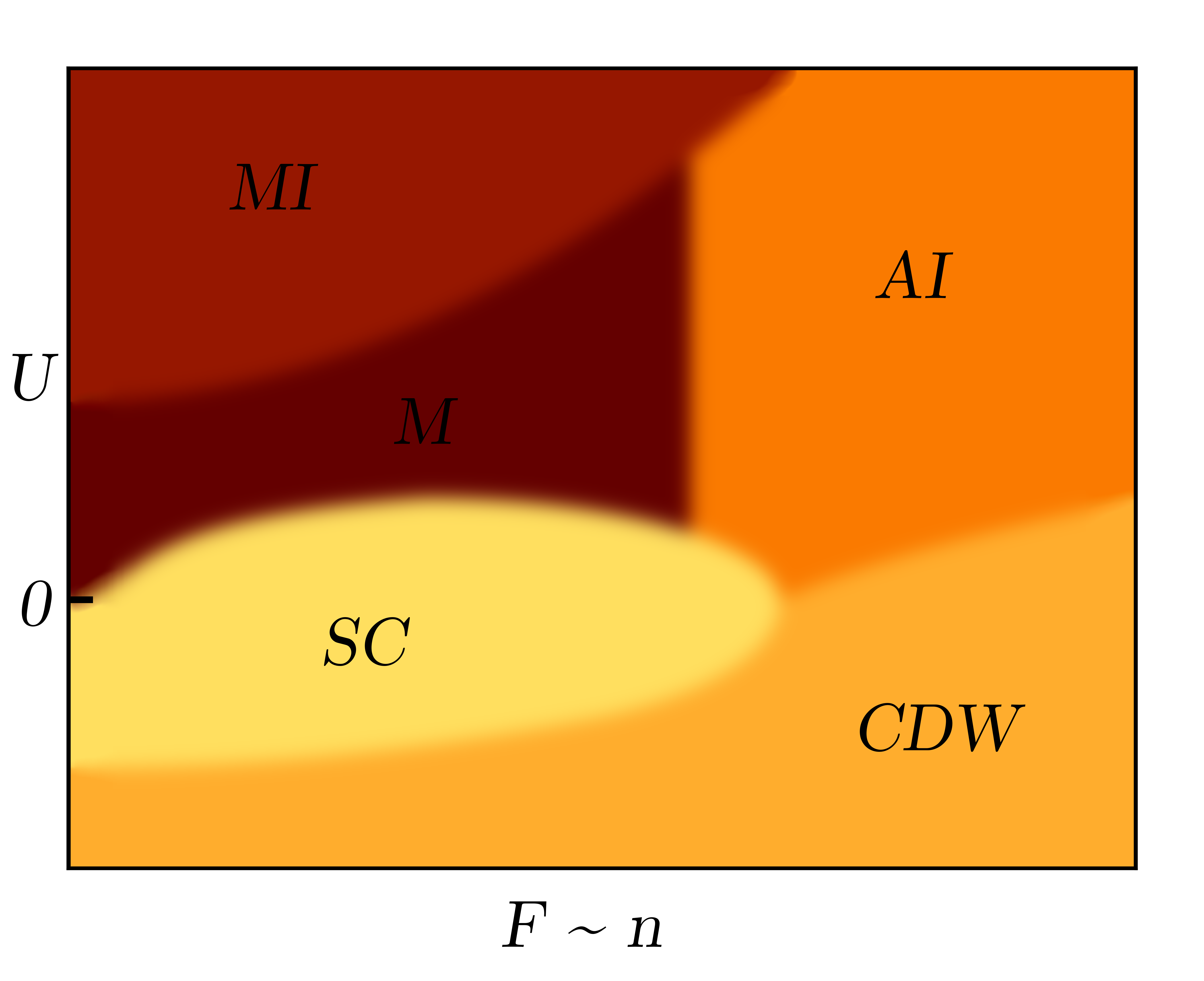} \caption{{\bf Schematic phase diagram of effective model, equation (3),} in the plane of bare interaction $U$ and pump fluence $F$ (equivalently, mean boson occupancy $n_{\rm B}$), assuming a half filled band.  Expected phases include a Mott Insulator (MI) phase occurring at large $U$ and small fluence,  metallic (M) and superconducting (SC) phases occurring at small to intermediate $U$ and small to intermediate fluence, and  Anderson/charge density wave insulator phases (AI/CDW) occurring at larger fluence.  }
\label{phase_diagram}
\end{figure}

Some aspects of the remarkably rich physics of equation~\eqref{Happrox} are summarized in the phase diagram shown in Fig.~\ref{phase_diagram}, sketched for simplicity for the case of the half-filled but non-nested band. $H_{\rm eff}$ is of course only an approximate description of a nonequilibrium situation. Phonon-assistend hopping will eventually equilibrate the electrons and the 'phases' shown indicate the qualitative characteristics expected of the intermediate time behavior as seen explicitly in our discussion of superconductivity below.  The on-site phonon number operator $\beta_{i}^{\dagger}\beta_{i}$ is a constant of the motion, so that in the absence of inelastic effects and phonon-phonon coupling (not included in equation~\ref{Happrox}) the phonon distribution is fixed by the pump field. For all but the last term in   $H_{\rm eff}$ we may replace the phonon number operator by its average over the phonon distribution function. The last term expresses the physics that different phonon occupancies on different sites  give rise to fluctuations of root mean square magnitude $g\sqrt{n_{\rm B}}$ in the on-site potential energy of the electrons, i.e. to an effective disorder. 

For simplicity we sketch the phase diagram for a half-filled non-nested band.  The pump pulse determines  the initial phonon density matrix and thus the mean value of the boson occupancy.    If the pump fluence is zero (only thermally excited phonons, occupation number $n_{\rm B}$ negligibly small at temperatures of order room temperature or less), then at large $U$ a Mott insulating phase occurs; as $U$ decreases the Mott insulator gives way to a metal and then at negative bare $U$ to a superconductor. As the fluence increases the effective $U$ decreases and the effective disorder also increases. The positive $U$ Mott insulating phase will cross over to an Anderson (disorder-dominated) insulator or a metal (depending on the bare value of $U$) and, when the $U^\star$ becomes negative enough, the material becomes a charge density wave  (interaction dominated) insulator. Of course it is important to emphasize again that the effective model is an approximate representation of a dynamical  physical situation: the phase represented here by the Anderson/CDW insulator is actually a slowly fluctuating polaronically trapped state.  Similarly if the initial interaction is not too negative (equilibrium metal), phonon excitation produces a superconductor, while if the initial interaction is attractive (equilibrium superconductor), then the strength of the superconductivity is increased.  Here it is important to note that s-wave superconductivity is robust to moderate disorder (Anderson theorem); only when the disorder becomes strong enough to localize the electrons will the superconducting state  give way to an Anderson/CDW phase. While induced superconductivity is a focus of current experimental attention,  the evolution of an initially Mott insulating regime will be of interest in the context of optically pumped Mott insulators.

We now present  physical consequences, focusing on the superconducting regime potentially relevant to recent  reports of transient superconductivity in phonon-pumped  $\mbox{K}_{3}\mbox{C}_{60}$. We note that while the simplified model introduced here omits many of the specifics of $\mbox{K}_{3}\mbox{C}_{60}$, including the three-fold orbital degeneracy of the electronic states and the multiplicity of on-ball phonons,  the basic physics of electrons subject to a phonon-mediated attractive interaction and a repulsion of electronic origin (non-negligible but in the end weaker than the phonon attraction) is agreed to describe the fullerides \cite{2003stronhang}. In mapping this material onto our simplified model we use a one band featureless semielliptical density of states (reasonably representative of typical densities of states of three dimensional materials) with a bare hopping parameter $J=35meV$  corresponding to the $\mbox{K}_{3}\mbox{C}_{60}$ bandwidth of approximately $0.42eV$ reported in \cite{2003stronhang}. We take the phonon frequency to be $\omega_0=0.17eV$, the frequency of the $T_{1u} (4)$ mode excited in the experiment and  assume that the combination of an initial laser pulse  and inelastic scattering places the phonons into a state described by a density matrix which is diagonal in the site and occupation number basis. The density matrix is characterized by a mean boson number $n_{\rm B}(t) $ with rise time and maximal value set by the experimental protocol of Mitrano {\em et al.} and a phenomenological  exponential decay representing  the physically present phonon decay processes not included in equation~\eqref{Happrox}.

For simplicity we treat the electronic physics within a BCS approximation using the  time dependent interaction determined by the phonon occupation via equation~\eqref{Ueff} and assume that the combination of bare electron-electron and linear electron-phonon coupling leads to a net attractive interaction producing superconductivity with an equilibrium transition temperature $T_c^{eq}=20K$. We also incorporated a weak phenomenological electronic damping to represent inelastic processes not well captured by the time-dependent BCS approximation. If this damping is not included the model maps onto an integrable model \cite{barankov2004collective} whose behavior, while of great theoretical interest, is substantially affected in practice by damping and by deviations from perfect integrability. It is interesting to note that underdamped gap oscillations were reported by Matsunaga \emph{et al.} \cite{matsunaga2013higgs} suggesting that the oscillations predicted by Barankov  \emph{et al.} may be observable. 

Estimates for the strength of the quadratic electron-phonon coupling $g$ are difficult to make.  Via the correlation between mode frequencies and doping, we use equation~\eqref{Happrox} along with experimental data to estimate $|g|$ to fall in the range $0.03-0.06$ (see Methods section).  As we shall see, producing effects at temperatures as high as those reported in the experiment requires a $g\sim 0.15$ within our model,  several times larger than that estimated from the phonon frequency shift, but even significantly smaller $g$ values would produce a sizable effect with an induced $T_c$ in the range of $50-100K$ (see Supplementary Information).  The crudeness of the estimation itself as well as the fact that our model is a simplified one-band, one mode model, suggests that it is reasonable to allow for a fair degree of latitude in choosing this value. Future work will be devoted to the study of a more realistic multi-band model to assess if a closer connection with the experimental setting can be obtained. 

\begin{figure}[h!]
	\centering{}\includegraphics[width=0.8\columnwidth]{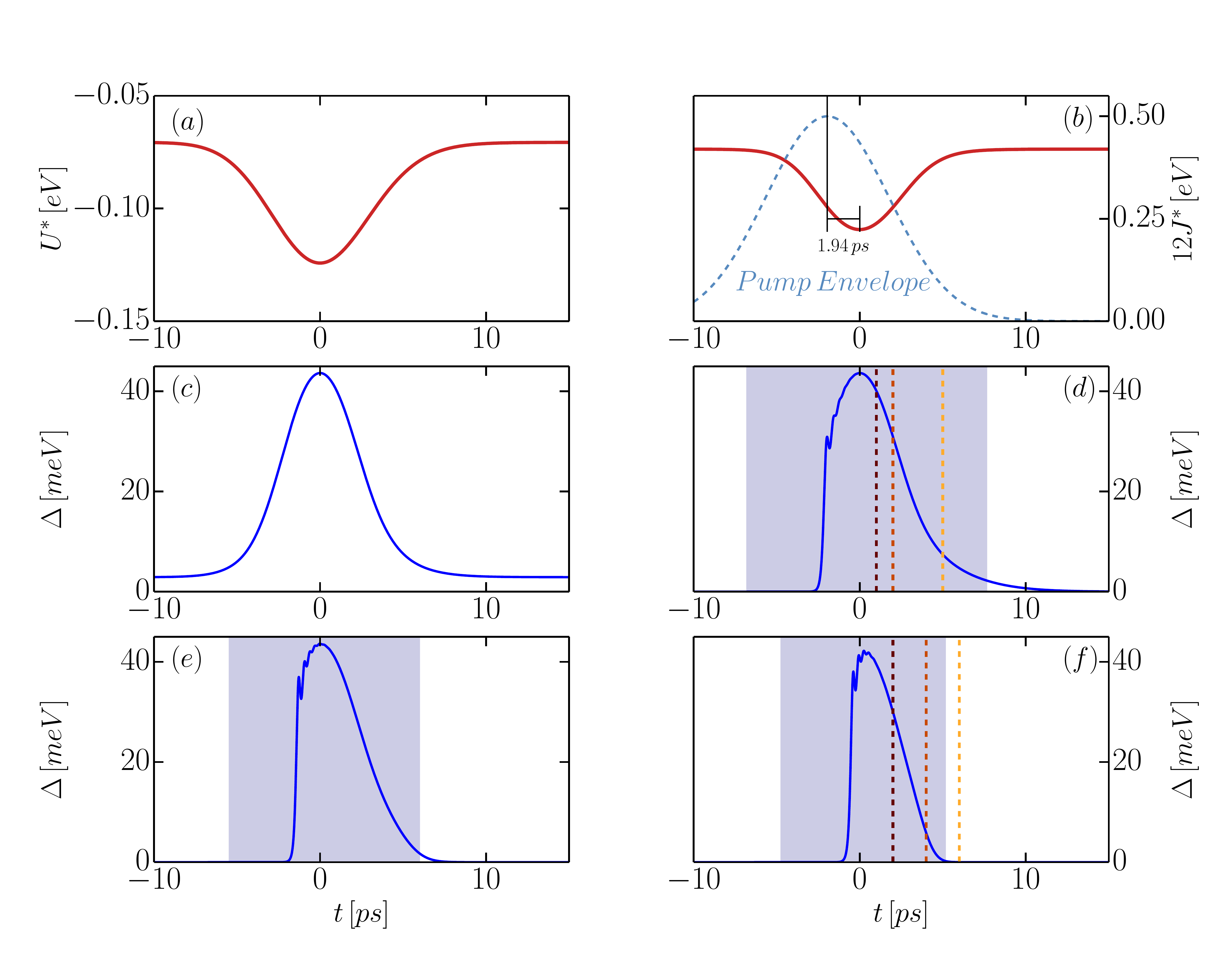}
	\caption{{\bf{Time dependence}} of the induced interaction (a), the renormalized hopping (solid line) and pump fluence profile (dashed line) (b) and the gap function (c-e) computed from equation~\ref{Happrox} as described in the methods section with  $U=-0.07eV$ (corresponding to an equilibrium superconducting critical temperature  $T_{\rm c}^{\rm eq}=20K$), $12J=0.42eV$, $\omega_0=0.17eV$, $g=0.15$, peak boson occupancy $n_B=14$   and  temperatures [$(c)$ $5K$, $(d)$ $25K$, $(e)$ $50K$ and $(f)$ $100K$].    For panels [$(d)-(f)$], the blue shaded area indicates the time window over which the induced interaction is strong enough to stabilize superconductivity at this temperature in equilibrium. The dashed horizontal lines indicate times at which the probe field is applied in the conductivity calculation (see Fig.~\ref{fig:Cond})  
}
	\label{fig:gapoft} 
\end{figure}

A rough estimate of the  maximum boson number $n_{\rm B}$ corresponding to the experimental situation can be made by comparing the  harmonic oscillator approximation for the relevant mode (here the $T_{1u}(4)$ mode) to the reported mode amplitude of $0.4$\AA\ $\sim 15-20\%$ of the equilibrium bond length, yielding $n_{\rm B} \sim 10-20$ per lattice site. It should be noted that for a realistic anharmonic mode this estimate may be conservative because the level spacing will decrease for larger $n_{\rm B}$ values.   As shown in the Supplementary Information, for  these values of $n_{\rm B}$ and the $g$ values we use the system is outside of, but close to, the disorder dominated region of the phase diagram.  In the following calculations we use the values  $n_{\rm B}=14$, $g=0.15$ and neglect the disorder.  The time dependence of the assumed pump field,  interaction and effective hopping  are shown in panels (a) and (b) of Fig.~\ref{fig:gapoft}. Note that the induced interaction and hopping renormalization are proportional to the time integral of the fluence so that the  maximum in the interaction amplitude lags the maximum in the  pump field by a time delay related to the pump profile width.

The remaining four panels of Fig.~\ref{fig:gapoft} show the time dependence of the superconducting gap amplitude, computed from the time-dependent BCS approximation using standard methods (see Supplementary Information).    We present results  for four representative temperatures. The lowest temperature ($T=5K<T_c$) illustrates that if the system is initially in the superconducting state, the gap amplitude is dramatically increased by the change in interaction, and follows the time dependence of the interaction closely. For the three cases in which the equilibrium material is in the normal state we see that  the gap is non-zero only over part of the pulse duration, with the time window where the gap is non-zero is decreasing as the temperature increases. The shaded regions  show the time regime in which the effective interaction is strong enough to sustain equilibrium superconductivity at the given temperature. We see that gap appears significantly after the dynamical interaction becomes strong enough to sustain equilibrium superconductivity.  The delay occurs because  the non-superconducting state is a marginally  unstable fixed point of the BCS approximation to the Hamiltonian, so the initial time dependence of the gap is controlled by an exponential amplification of small fluctuations. 

\begin{figure}[t!]
\centering{}\ \includegraphics[width=0.8\columnwidth]{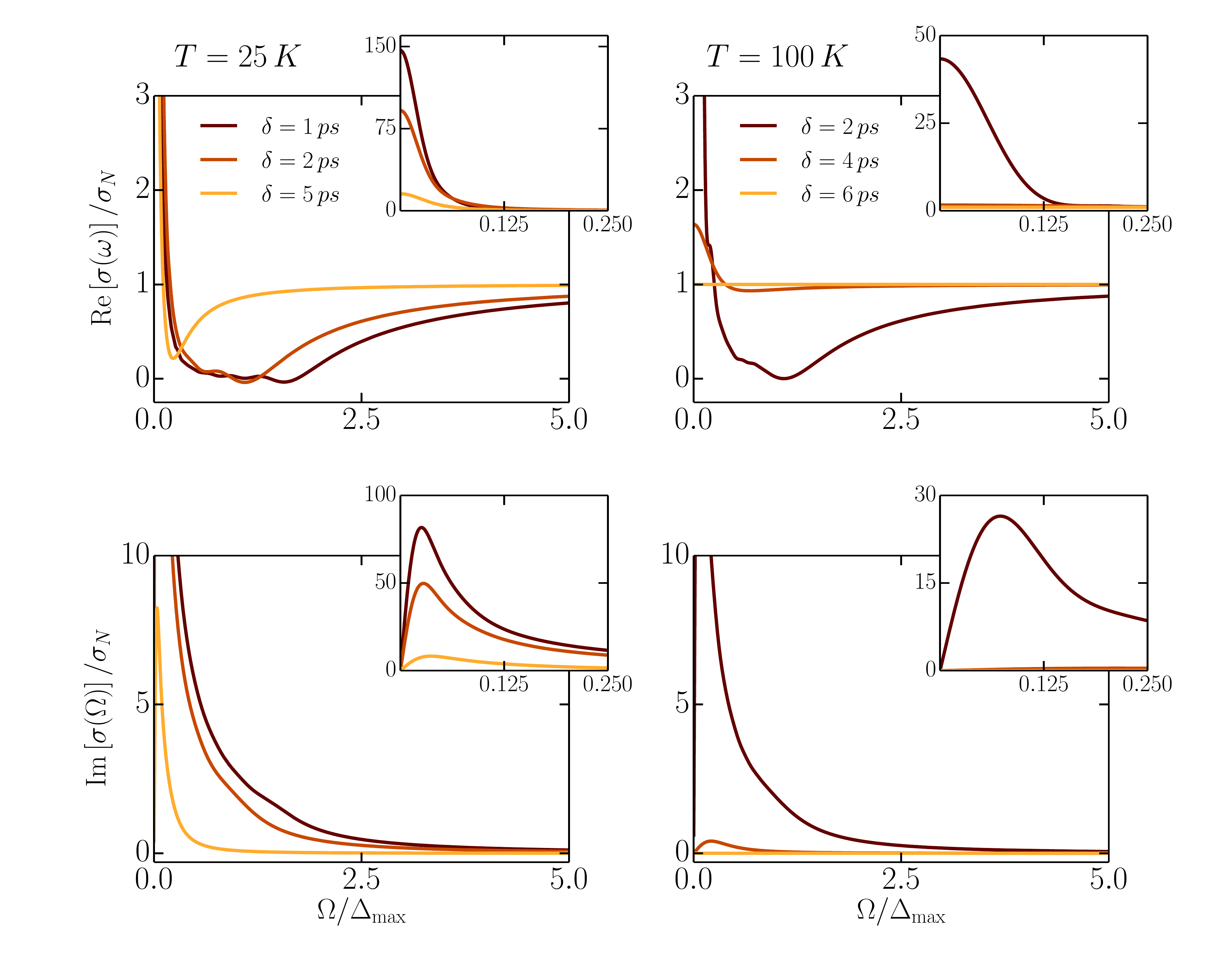}
\caption{{\bf{Real  and imaginary parts of non-equilibrium optical conductivity}} defined as the response to a delta-pulse probe field applied at different times relative to the maximum of the induced interaction  and calculated for temperatures $T=25K$ (left column) and $T=100K$ (right colum) using  the gap profiles $\Delta(t)$ of Fig.~\ref{fig:gapoft}. The x-axis is scaled with respect to the maximum value of the corresponding gap profiles shown in Fig.~\ref{fig:gapoft}.  The insets show $\rm Re\sigma\left(\omega\right)\left[\rm Im\sigma\left(\omega\right)\right]$ for a larger y-scale and a zoom of the low $\Omega$ scale. }
\label{fig:Cond} 
\end{figure}

The optical response is an important probe of transient superconductivity. A complete discussion presents subtle experimental and theoretical issues which are beyond the scope of this paper. Here we {\em define} the nonequilibrium conductivity $\sigma(t,t^\prime)$ via the relation $j(t)=\int dt^\prime\sigma(t,t^\prime)E_{\rm probe}(t^\prime)$ connecting the time dependent current $j(t)$ to the time dependent amplitude of a probe electric field $E_{\rm  probe}\left(t\right)$. This definition reduces to the familiar result at equilibrium and corresponds approximately to the quantity presented in the experimental literature.  We assume for simplity that the probe pulse has negligible width, an integrated amplitude $E_P$ and is applied after a delay $\delta$ defined relative to the maximum in the induced interaction (here set to be $t=0$; note that this occurs somewhat after the maximum in the pump field). We Fourier transform $j(t)$ and define a frequency-dependent nonequilibrium conductivity $\sigma(\Omega)$ as the ratio of $j(\Omega)$ to $E_P$. We compute the conductivity using an adiabatic approximation and the dirty limit  (Mattis-Bardeen \cite{mattis1958theory}) formula (details are given in the Methods section).

Representative results are shown in Fig.~\ref{fig:Cond}. The real part of the conductivity displays a gap-like structure that depends on temperature and on the delay $\delta$ between the pump that establishes the transient superconducting state and the probe electric field. The imaginary part correspondingly displays an intermediate frequency enhancement. The subgap conductivity displays oscillations arising physically from the transient nature of the superconducting state;  the real part may even become slightly negative.  For frequencies below a scale $\Omega_{\rm low}$ that depends on the lifetime of the superconducting state the system behaves like a normal conductor, with the imaginary part of the conductivity vanishing linearly and the real part saturating.   The results, including the subgap oscillations and  variation with  delay between the pump and probe  pulses,  bear a striking similarity to the measurements reported in \cite{mitrano2016possible} (the $\Omega<\Omega_{low}$ regime has not yet been experimentally resolved).  

In summary, we have introduced a robust and general mechanism by which a large amplitude coherent phonon excitation can dramatically change the electronic behavior of a system. The effect is generically symmetry-allowed and is the dominant coupling mechanism in centro-symmetric structures. The main requirements are an optically active phonon of a reasonably high frequency with a frequency that depends modestly   on the electron concentration ($\sim 5\%-10\%$ over the range $0\leq n\leq 2$).  Under these circumstances a high degree of phonon excitation produces a  component to the electron-electron interaction which is always attractive, is proportional to the number of excited phonons and to the bare phonon frequency and for reasonable couplings  may be large enough to appreciably affect electronic properties. We have shown here that for reasonable parameters the mechanism can produce transient high temperature superconductivity, but the applications are expected to extend far beyond this case. For example, in  transition metal oxides or dicalcogenides, the dependence of breathing-mode and Jahn-Teller phonons on d-shell occupancy implies an excitation-induced negative-$U$  that has the potential to tune systems across the Mott metal-insulator or charge density wave transition. The application of these ideas to enhanced superconductivity in the cuprates is an important open problem, made more difficult by the lack of consensus on the pairing mechanism (and indeed the important physics) in these systems. If the materials are doped Mott insulators, the reduction of U following from the mechanism proposed here could enhance the superexchange $J\sim t^2/U$ thereby increasing the pairing, however a detailed treatment is beyond the scope of this paper

The results presented here raise many questions now under investigation, including the effects of phonon bandwidth and anharmonicity on the evolution of phonon coherence during and after the initial phonon pulse, the consequences of electronic relaxation and the dynamical aspects of the electron-phonon coupling, the implications of electronic orbital degeneracy and multiple phonon modes, and the physics of the quasilocalization induced at larger fluence.  Realistic {\em ab-initio} calculations of the crucial coupling constant $g$ in different materials are also of interest. 

\hspace{0.5in}

\noindent\emph{Acknowledgements} AJM and EYW were supported by the Basic Energy Sciences Program of the US Department of Energy under grant SC-0012592. DMK was supported by DFG KE 2115/1-1. DRR was supported by NSF CHE-1464802.  \\
\emph{Author contributions} All authors contributed to planning the research, developing the methods interpreting the results and writing the paper. EYW and DMK performed the numerical calculations. \\
\emph{Competing financial interests:}
The authors declare no competing financial interests.

\newpage

{\bf Methods} 

{\it Estimation of dimensionless coupling:}  The parameter $g$ is estimated from the density dependence of the phonon frequency via 
\begin{equation}
\omega_{0}(\hat{n}_{i})\equiv\omega_{0}\left(1+2g\left(\hat{n}_{i\uparrow}+\hat{n}_{i\downarrow}\right)\right)^{\frac{1}{2}}\label{eq:freq_shift}
\end{equation}
with the bare phonon frequency $\omega_{0}=\sqrt{K/M}$. It should be noted that this is a crude estimate since experimental doping also induces structural changes, and our model is highly simplified.

Measured vibrational spectra show a frequency shift of $\Delta\omega=-7.56,-1.49,0,-10.66\,\,meV$ for the non-symmetric $T_{1u}\left(1-4\right)$ respectively for $A_{x}C_{60}$ with $x=6$ \cite{giannozzi1996effects}. Using our prediction for the frequency shift in equation~\eqref{eq:freq_shift} we find $\left|g\right|\approx0.03-0.06$ for non-zero shift values when we associate $x=6$ (filled three bands) to $\hat{n}_{i\uparrow}+\hat{n}_{i\downarrow}=2$ (filled single band). We have also examined the case $x=3$ with similar results. This estimated value of $|g|$ is $2.5-4.5$ times smaller then the value used in the main text $g=0.15$, for which our theory stabilizes superconductivity on comparable scales as found in the experiment. Even lower $g$  values produce sizable effects.  

{\it Time dependence of boson occupancy} In the experiment the phonons are excited via a light pulse with the approximate temporal profile
\begin{equation}
F_d\left(t\right)=E_0e^{-\frac{t^{2}}{2\sigma^{2}}}\sin\Omega t,\label{eq:light_field}
\end{equation}
with $E_0$ determined from the  fluence of $1.1\,{\rm mJ\,cm^{-2}}$ given by Mitrano et al \cite{mitrano2016possible}, mean frequency $\Omega$ tuned to the C$_{60}$ $T_{1u}$ optical phonon frequency ($\omega_0$ in the notation of the present paper) and pulse width $\sigma$ such that the envelope corresponds to about 50 cycles of the phonon. In our calculations we used $\sigma=48\pi/\omega_0$. 

The light field is, to good approximation laterally uniform across the sample and couples linearly to the mode amplitude (the decay across the sample due to the finite optical penetration depth is included in the estimate of the fluence given by Mitrano et al). In the absence of electron-phonon coupling textbook quantum mechanics shows that the resulting phonon state would be a coherent state with maximum oscillator amplitude $x_{\rm max}$ determined by:
\begin{equation}
x_{\rm max}\left(t\right)=\sqrt\frac{m\omega_0}{2\hbar}\left|\int_{0}^{t}e^{-i(\omega_0+i\gamma_{\rm ph})\left(t-t'\right)}F_d\left(t'\right)\mbox{d}t'\right|.\label{eq:respond_function_to_light}
\end{equation}
Here we have included a phenomenological damping $\gamma_{\rm ph}=0.01\Omega$ to account for the leakage of energy out of the phonon modes.   Using the textbook  relation between oscillator amplitude and coherent state and setting $M$ equal to the mass of a $C_{12}$  atom  we  estimate a maximum of $n_{\rm B}< 20 $ in the experiment of Ref.~\cite{mitrano2016possible} per lattice site from the reported mode amplitude of $0.4$\AA\ $\sim 15-20\%$ of the equilibrium bond length.  Because the electron-phonon coupling destroys the coherence between different occupation number states on a time scale comparable to the pump pulse width we approximate the phonon state at all times in terms of a density matrix which is diagonal in the phonon occupation number basis with occupation probabilities corresponding to the projection of the  coherent state with oscillation amplitude $x_{\rm max}(t)$ onto an occupation number diagonal density matrix. 

{\it Transient superconductivity:}

We treat equation~\eqref{Happrox} in a time-dependent  BCS approximation and neglect disorder effects, so the retarded (=0 for $t<t^{\prime}$) and advanced (=0 for $t>t^{\prime}$) electrons Green's functions are obtained from  (we do not denote the spatial indices explicitly)
\begin{eqnarray}
i\partial_{t}\mathbf{G}^{R}(t,t^{\prime}) & = & \mathbf{H}(t)\mathbf{G}^{R}(t,t^{\prime}),\label{GR}\\
-i\partial_{t}\mathbf{G}^{A}(t^{\prime},t) & = & \mathbf{G}^{A}(t^{\prime},t)\mathbf{H}(t),\label{GA}
\end{eqnarray}
where $\mathbf{H}$ and $\mathbf{G}$ are matrices in Nambu space.  Within the time dependence BCS approximation the electron distribution does not relax  and the Keldysh function is  
\begin{equation}
\mathbf{G}^{K}(t,t^{\prime})  =  \mathbf{G}^{R}(t,0)\left(\mathbf{1}+2f^{0}_k\tau_{3}\right)\mathbf{G}^{A}(0,t^{\prime})
,\label{GKdef}
\end{equation}
with initial distribution parametrized by $f_0$, which we typically take to be the Fermi-Dirac distribution (non=superconducting initial state). For a  superconducting initial state we choose  $f^{0}_k \mathcal{M}^\dagger\tau_{3} \mathcal{M}$, where $ \mathcal{M}$ are the rotations that diagonalize the initial Hamiltonian in Nambu-space and in $f_0$ we use the superconducting energies.   

The time-dependent Hamiltonian is  
\begin{equation}
\mathbf{H}(t)=\left(\begin{array}{cc}
\varepsilon_{k}\left(t\right) & \Delta(t)\\
\Delta^{*}(t) & -\varepsilon_{k}\left(t\right)
\end{array}\right),\label{Hoft}
\end{equation}
where the time dependence of the gap function $\Delta(t)$ is computed from \begin{equation}
\Delta(t)=U(t)\sum_{k}Tr\left[\tau^{-}\mathbf{G}^{K}(t,t)\right],
\end{equation}
with the time dependence of $\varepsilon_k$ and $U$ given by the time-dependent phonon distribution.  

We write the time-dependent gap equation as
\begin{equation}
\partial_{t}\vec{\Omega}=2\vec{b}_{\rm eff}\times\vec{\Omega},
\label{gapeqfinal}
\end{equation}
where 
\begin{equation}
\vec{b}_{\rm eff}\left(t\right)=\varepsilon_{k}\left(t\right)\hat{e}_{3}+{\rm Re}[\Delta\left(t\right)]\hat{e}_{1}-{\rm Im}[\Delta\left(t\right)]\hat{e}_{2}.
\label{beffdef}
\end{equation}

We solve equation~\eqref{gapeqfinal} by propagating forward in finite time steps $\Delta t/J=0.056$. We additionally introduce a phenomenological decay to represent electronic dephasing and energy loss processes by  taking at each time step the weighted average of 99\% of the equal time Keldysh Green's function obtained by equation~\eqref{GKdef}  and  1\% of the instantaneous equilibrium Keldysh Green's function. This  weighting  produces a decay towards equilibrium with relaxation rate $\gamma=0.15J$. 

Essentially, equation~\eqref{gapeqfinal} was written down and solved by  Barankov {\em et al.} \cite{barankov2004collective} for an undamped integrable system and a piecewise constant $U(t)$. Our solution is slightly different  because we employ the $U(t)$ implied by the experimental pump pulse and the phenomenological decay and include electronic damping as described above. 

{\it Optical conductivity:}
It is beyond the scope of this paper to model the  details of pump-probe nonequilibrium conductivity measurements. For the purposes of this paper we assume that a weak applied (`probe') electric field  $E_{probe}(t^\prime)$ produces an electric current $j(t)$ which is non-vanishing only at $t>t^\prime$ (recall that the drive pulse excites a zone-center optical phonon which does not directly produce an electronic current). The relationship $j(t)=\int dt^\prime \sigma(t,t^\prime)E(t^\prime)$  between  the current and the strength $E_0$ of the probe field then {\em defines} the  conductivity.

We introduce the center time $T=(t+t^\prime)/2$ and relative time
$t_{rel}=t-t^\prime$ and Fourier transform the defining relationship to obtain 
\begin{equation}
j(\Omega)=\int dTdt_{rel}e^{-i\Omega\left(T+\frac{t_{rel}}{2}\right)}\sigma(T,t_{rel})E_{\rm probe}\left(T-\frac{t_{rel}}{2}\right).\label{linrespE2}
\end{equation}

We further suppose that the pump pulse is short on time-scales of interest, and is peaked at a delay time $t_D$ relative to the time at which the interaction $U(t)$ is maximal
\begin{equation}
E_{\rm probe}(t)=E_P\delta(t-t_D)
\label{Eprobedef}
\end{equation}
so that $t_{rel}=2T-2t_D$ and 
\begin{equation}
\frac{j(\Omega)}{E_P}=\int dTe^{-i\Omega\left(2T-t_D\right)}\sigma(T,2T-2t_D)
\label{linrespE3}
\end{equation}
{\em defining} $\sigma(\Omega;t_D)$ as the right hand side of equation~\eqref{linrespE3}.

We approximate the conductivity $\sigma(T,2T-2t_D)$ as the equilibrium conductivity appropriate to the gap at the average time, i.e.
\begin{equation}
\sigma(T,t_{rel})\approx\sigma_{equil}^{\Delta(T)}(t_{rel})
\label{adiabaticsigma}
\end{equation}
so that equation~\eqref{linrespE3} becomes
\begin{equation}
\frac{j(\Omega)}{E_P}=\int_{t_D}^\infty dTe^{-i\Omega\left(2T-t_D\right)}\sigma_{eq}^{\Delta(T)}(2T-2t_D).
\label{linrespadiabatic}
\end{equation}

We evaluate $\sigma_{eq}$ using the Fourier transform of the dirty-limit (Mattis-Bardeen) formula (equations (3.9) and (3.10) of \cite{mattis1958theory}). The computation of the  Fourier transform of the dirty-limit conductivity is simplified by the observation that the superconducting contribution is a universal function of $\Omega/\Delta$ and $T/\Delta$  as well as by adding and subtracting the normal state conductivity. 
\newpage
\begin{center}
{\large \bf{Transient superconductivity from electronic squeezing of optically pumped phonons - Supplementary Information}}

\end{center}

\section{Derivation of Effective Model} 

This section presents the details of the derivation of the effective model, equation (3) of the main text,  from the fundamental model, equation (1) of the main text.

The exact Hamiltonian may be written (here we neglect the bare electron-electron repulsion for simplicity and denote the original boson operators by $b$):
\begin{equation}
{H}_{\rm{exact}}=-\sum_{ij\sigma}J_{ij}c_{i\sigma}^{\dagger}c_{j\sigma}+\omega_{0}\sum_{i}\left(b_{i}^{\dagger}b_{i}+\frac{1}{2}\right)+\frac{g\omega_{0}}{2}\sum_{i}\left(b_{i}^{\dagger}+b_{i}\right)^{2}\hat{n}_i,\label{eq:exact}
\end{equation}
with on-site electron density operator $\hat{n}_i=\sum_{\sigma}c_{i\sigma}^{\dagger}c_{i\sigma}$.

We treat the quadratic electron-phonon coupling (third term of equation~\eqref{eq:exact}) via a squeezing transformation $H\rightarrow H_{squeezed}= e^{\hat{S}}He^{-\hat{S}}$ with $\hat{S}=\frac{i}{2}\sum_{j}\zeta_{j}\left(\hat{x}_{j}\hat{p}_{j}+\hat{p}_{j}\hat{x}_{j}\right)$ where $\hat{x},\hat{p}$ are the phonon displacement and momentum operators  and
\begin{equation}
\zeta_{j}=-\frac{1}{4}\ln\left[1+2g\hat{n}_{j}\right].
\label{zetadef}
\end{equation}

Denoting the operator creating a squeezed phonon state on site $i$ as $\beta^\dagger_i$ we obtain
\begin{equation}
H_{\rm{squeezed}}=-e^{\hat{S}}\sum_{ij\sigma}J_{ij}c_{i\sigma}^{\dagger}c_{j\sigma}e^{-\hat{S}}+\sum_{i}\omega\left[\hat{n}_i\right]\left(\beta_{i}^{\dagger}\beta_{i}+\frac{1}{2}\right),
\label{eq:squeezed}
\end{equation}
with
\begin{equation}
\hat{S}=\sum_{j}\zeta_{j}\frac{\beta_j^\dagger \beta_j^\dagger-\beta_j\beta_j}{2},
\label{Sbeta}
\end{equation}
and
\begin{equation}
\omega\left[\hat{n}\right]=\omega_0\sqrt{1+2g\hat{n}}.
\label{omegaofn}
\end{equation}

We begin an analysis of equation~\eqref{eq:squeezed} with the second term, which may be expanded in powers of $g$ which is small in the situations of interest here.  Expanding equation~\eqref{omegaofn} to $\mathcal{O}(g^2)$, using  $\hat{n}=n_\uparrow +n_\downarrow$, noting that $n_\sigma^2=n_\sigma$  and inserting the result into equation~\eqref{eq:squeezed} we obtain three terms: 
\begin{equation}
H_{ph}=\sum_i\left(\omega_0\left(\beta_{i}^{\dagger}\beta_{i}+\frac{1}{2}\right)+\omega_0\left(g-\frac{g^2}{2}\right)\left(\beta_{i}^{\dagger}\beta_{i}+\frac{1}{2}\right)\hat{n}_i-g^2\omega_0\sum_i\left(\beta_{i}^{\dagger}\beta_{i}+\frac{1}{2}\right)\hat{n}_{i\uparrow}\hat{n}_{i\downarrow}\right).
\label{interactions}
\end{equation}
The first term is the bare phonon Hamiltonian, expressed in terms of the squeezed states, the second term is a phonon state-dependent potential  felt by the electrons, and the third term is the phonon-induced electron-electron interaction, which as noted is attractive. We replace the phonon operators in the last term by their expectation value and discuss the middle term below. 

We now consider the transformation of the hopping ($J$) term. Noting that operators on different sites commute and writing the hopping term in the electron occupation number basis we obtain (here the $n_i$ refer to the occupation numbers in the initial states before hopping  and $\mathcal{O}_j=\frac{\beta_j^\dagger \beta_j^\dagger-\beta_j\beta_j}{2}$)
\begin{equation}
\tilde{H}_{hop}\left[\{n_i,n_j\}\right]=-\sum_{ij\sigma}J_{ij}
c^\dagger_{i\sigma}c_{j\sigma}e^{i\left(\zeta[n_i+1]-\zeta[n_i]\right)
\mathcal{O}_i}e^{i\left(\zeta[n_j-1]-\zeta[n_j]\right)\mathcal{O}_j}.
\label{Htilde1}
\end{equation}
The phonon creation/annihilation operator content of $\hat{S}$ means that hopping processes may involve the creation or annihilation of pairs of phonons. As is commonly done in the linearly coupled polaron problem we approximate the hopping by its expectation value (the adequacy of this approximation is addressed in the next section).   Computing the expectation value exactly is awkward because the commutator $\left[\beta^\dagger\beta^\dagger,\beta\beta\right] $ does not commute with $\beta\beta$ or $\beta^\dagger\beta^\dagger$ so  the exponentials  cannot easily be disentangled  into the product of an exponential of creation operators times an exponential of annihilation operators times a c-number.    

We now expand the exponentials, noting that  $\zeta[n_i\pm1]-\zeta[n_i]= \pm \frac{g}{2}+\mathcal{O}(g^2\hat{n})$ and retaining only terms that give a non-vanishing expectation value in the decohered (phonon density matrix diagonal in occupation number basis) state. The leading term with a non-zero expectation value is then $-\frac{g^2}{32}\left<\beta^\dagger_i\beta^\dagger_i\beta_i\beta_i+\beta_i\beta_i\beta^\dagger_i\beta^\dagger_i+\beta^\dagger_j\beta^\dagger_j\beta_j\beta_j+\beta_j\beta_j\beta^\dagger_j\beta^\dagger_j\right>$. Evaluating the expectation value of this term in the decohered state and assuming all sites are the same on average and re-exponentiating we obtain 
\begin{equation}
J^\star=e^{-\frac{g^2}{8}\left<n^2+n+1\right>}J=e^{-\frac{g^2}{8}\left(n_{\rm B}^2+2n_{\rm B}+1\right)}J.
\label{Jstar}
\end{equation}
where in the second equality we assumed that the occupation number eigenvalues are Poisson-distributed with mean occupancy $n_{\rm B}$.   We see that the expansion parameter is $gn_{\rm B}$ times a favorable numerical factor. 

\section{Two site model: exact and approximate Hamiltonian}.  

\subsection{Overview; Exact and Approximate  Hamiltonians; Calculational Procedure}

In order to gain insight into the physics of the model  and the limits of validity of the approximations leading to the effective model, in this section we compare the results of  numerically exact solutions of the two site version of the fundamental (equation (1) of the main text) and approximate (equation (3) of main text) Hamiltonians. 

The two-site version of the exact Hamiltonian is

\begin{equation}
{H}_{\rm{exact}}^{2-site}=-J\sum_{\sigma}\left(c_{2\sigma}^{\dagger}c_{1\sigma}+H.c.\right)+\omega_{0}\sum_{i=1,2}\left(b_{i}^{\dagger}b_{i}+\frac{1}{2}\right)+\frac{g\omega_{0}}{2}\sum_{i=1,2}\left(b_{i}^{\dagger}+b_{i}\right)^{2}\hat{n}_i,
\label{eq:exact_2site}
\end{equation}

while the two-site version of the approximate Hamiltonian  is 
\begin{eqnarray}
{\tilde{H}}_{\rm{eff}}^{2-site}& = & -J^\star \sum_{\sigma}\left(c_{2\sigma}^{\dagger}c_{1\sigma}+H.c.\right)+\omega_{0}\sum_{i=1,2}\left(\beta_{i}^{\dagger}\beta_{i}+\frac{1}{2}\right)\nonumber \\
 & + & \omega_{0}\left(\frac{g}{2}-\frac{g^{2}}{4}\right)\sum_{i\sigma}\left(2\beta_{i}^{\dagger}\beta_{i}+1\right)
 n_{i\sigma}+U^\star\sum_{i=1,2}n_{i\uparrow}n_{i\downarrow},
\label{eq:approx_2site}
\end{eqnarray}
with $J^\star$ given by equation~\eqref{Jstar} and $U^\star=-\frac{\omega_{0}g^{2}}{2}\left(2n_{\rm B}+1\right)$.

In the absence of electron-phonon coupling the pump pulse used  in the experiment of Mitrano {\em et al.} \cite{mitrano2016possible} would produce an equally phased  oscillator coherent state on each site. We therefore integrate equations~\eqref{eq:exact_2site} and \eqref{eq:approx_2site}  forward in time  from an initial state taken to be the direct product of the noninteracting ($g=0$) two electron ground state and a phonon coherent state of mean occupancy $n_{\rm B}=9$ and identical phase  on each site and compute both time-dependent and time-averaged electron and phonon properties. (As long as $g^2<1$ and  $n_{\rm B}\gg 1$ the comparison is insensitive to the precise values of $g$ and $n_{\rm B}$). Solving the exact model requires truncating the boson Hilbert space. We increased the number of states until convergence on the respective plots is reached  \cite{Clark}. In all of the plots presented in this section the unit of time is $1/J$. Using the estimate $J=35meV$ means that $t=100/J$ corresponds to $3.3ps$.

\subsection{Phonon Properties}

\subsubsection{Coherence}

\begin{figure}[t]
\includegraphics[width=\columnwidth]{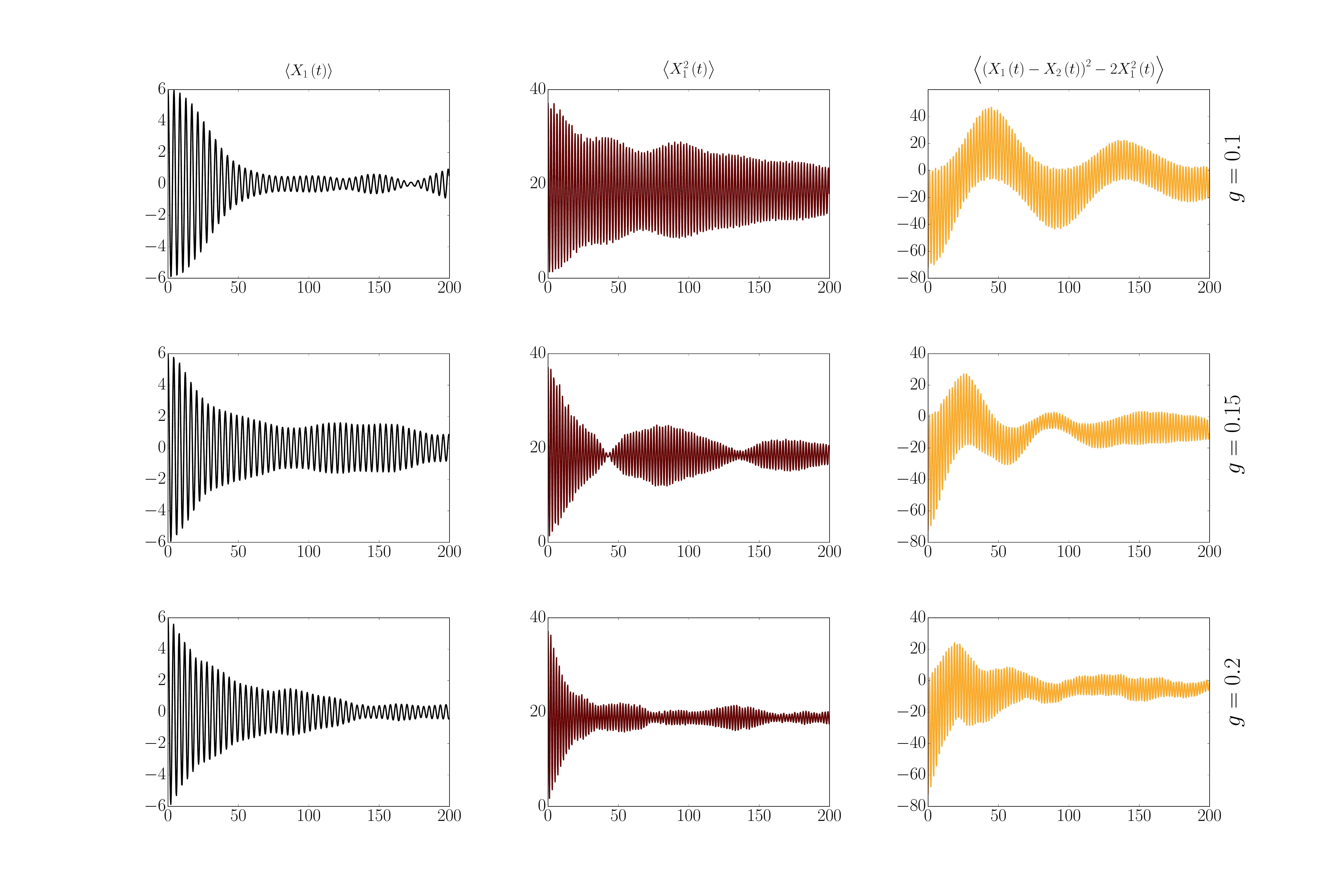}
\caption{{\bf Coherence properties of oscillator states} computed from equation~\eqref{eq:exact_2site} with $\omega_0=1.36J$, starting from coherent states with boson expectation value $n_B=9$  and non-interacting electron ground state for electron-boson coupling $g=0.1$ (upper panels), $g=0.15$ (middle panels) and $g=0.2$ (lower panels). Left panels: time dependence of expectation value of position coordinate of the oscillator on site 1. Middle panels, time dependence of expectation value of square of position coordinate of the oscillator on site 1. Right panels, left-right correlation function $\left<\left(X_1(t)-X_2(t)\right)^2-X_1^2(t)-X_2^2(t)\right>$. We obtained the results with 40 boson states on each site.   }
\label{fig:coherence}
\end{figure}

We begin by  using the exact two-site model to study the evolution of the oscillator states from the initial condition of equally phased coherence states. The left panel of Fig.~\ref{fig:coherence} shows the time dependence of the expectation value of the position coordinate of one of the two oscillators (the other is identical). The curves display the expected free-boson coherent state oscillations $X(t)=X_0\cos(\omega\left[n\right] t)$ but the oscillations decay within about 15 cycles, showing that coherence in the initial state is rapidly lost.  The central panel shows $X_1^2(t)$; we see that the oscillator state converges to one with small fluctuations about a reasonably well defined mean oscillation level, indicating that the oscillators remain in a relatively highly excited state described to good approximation by a time independent distribution with negligible phase coherence between different number states.  Finally the right panels show that the intersite correlation, which is perfect in the initial state, decays to a small (albeit non-zero) value, about $10\%$ of the mean value of $X_1(t)^2$. 

The electronic timescales relevant to superconductivity are of the order of hundreds to thousands of $J^{-1}$, so these plots justify the  neglect of phonon coherence (either between sites or between different occupation numbers on site) in our approximate model. 

\subsubsection{Phonon number state occupations}

The results of the previous subsection establish that the initial phonon coherence decays rapidly, so that in the phonon occupation number basis the density matrix describing the oscillators on one site is approximately diagonal. In this subsection we examine the time dependence of the diagonal elements of the density matrix in more detail. The key issue is that the approximate model conserves phonon number on each site, so in the approximate model an initially diagonal (in the occupation number basis) density matrix does not change with time.

\begin{figure}[t]
\includegraphics[scale=0.22]{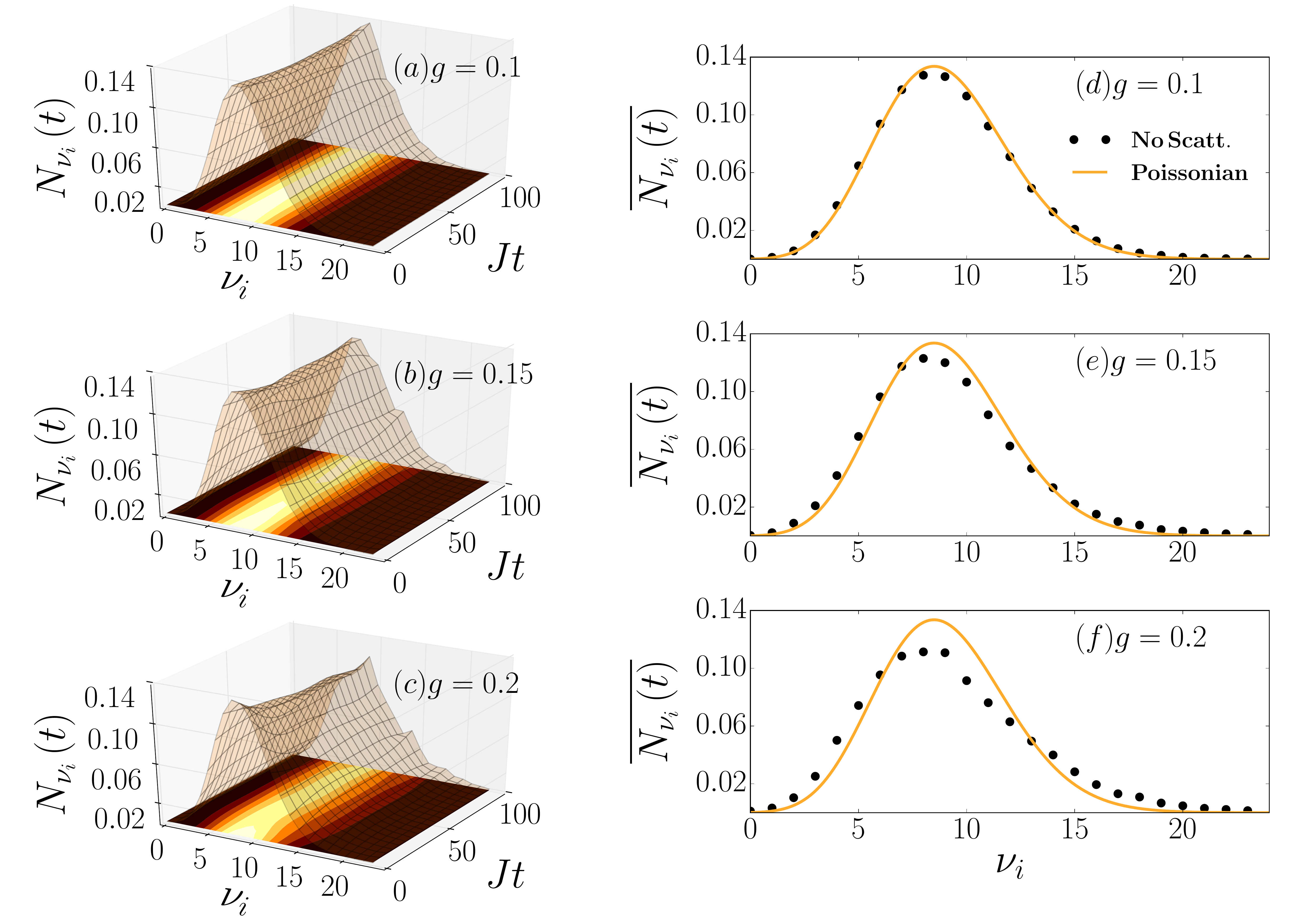}
\caption{{\bf{two-site model comparison. }} Panels $\left(a-c\right)$ show the time dependence of the  probability ${N_{\nu_i}}\left(t\right)$ of measuring the number eigenstate $\nu_i$ computed using  the exact two-site polaronic model equation~\eqref{eq:exact_2site} with initial state a coherent state with $n_{\rm B}=9$ on each site and $\omega_0=1.36J$.  Panels $\left(d-f\right)$ show the time average of  $\overline{{N_{\nu_i}}\left(t\right)}$ (dots) compared with the Poisson distribution (line) characterizing the initial state. We obtained the results with 30 boson states on each site. }
\label{two_site_comp_phonons}
\end{figure}

In the initial coherent state the occupation probabilities (diagonal density matrix elements) are Poisson-distributed; if the mean boson occupancy is $n_{\rm B}$ then the occupation probabilities $N_\nu$  of the $\nu$-phonon occupation number eigenstates are $N_\nu=e^{-n_B}n_{\rm B}^\nu/\nu!$.  The left hand panels  $\left(a-c\right)$ of Fig.~\ref{two_site_comp_phonons} present the time dependence of  $N_\nu$ obtained by solving the exact model  for different values of the dimensionless electron-phonon coupling $g$ for $\omega_{0}=1.36J$. We see that the diagonal matrix elements of the initial density matrix  are to a good approximation preserved over times long compared to the basic electronic hopping scale. A different perspective on these results is shown in panels   $(d-f)$, which compare the time averaged $\overline{N_{\nu_i}(t)}$ to the Poisson distribution corresponding to the initial state (which is preserved under the dynamics defined by the approximate model). These results show that for the values of dimensionless coupling $g$  and mean boson occupancy $n_B$ relevant here, the changes of the diagonal density matrix elements from the initial Poisson distribution values are not important: approximation of the phonon distribution as an incoherent average over Poisson-distributed occupation-number eigenstates is reasonable. 

\subsection{Electronic properties}

\subsubsection{Double Occupancy}

\begin{figure}[t]
\includegraphics[width=\columnwidth]{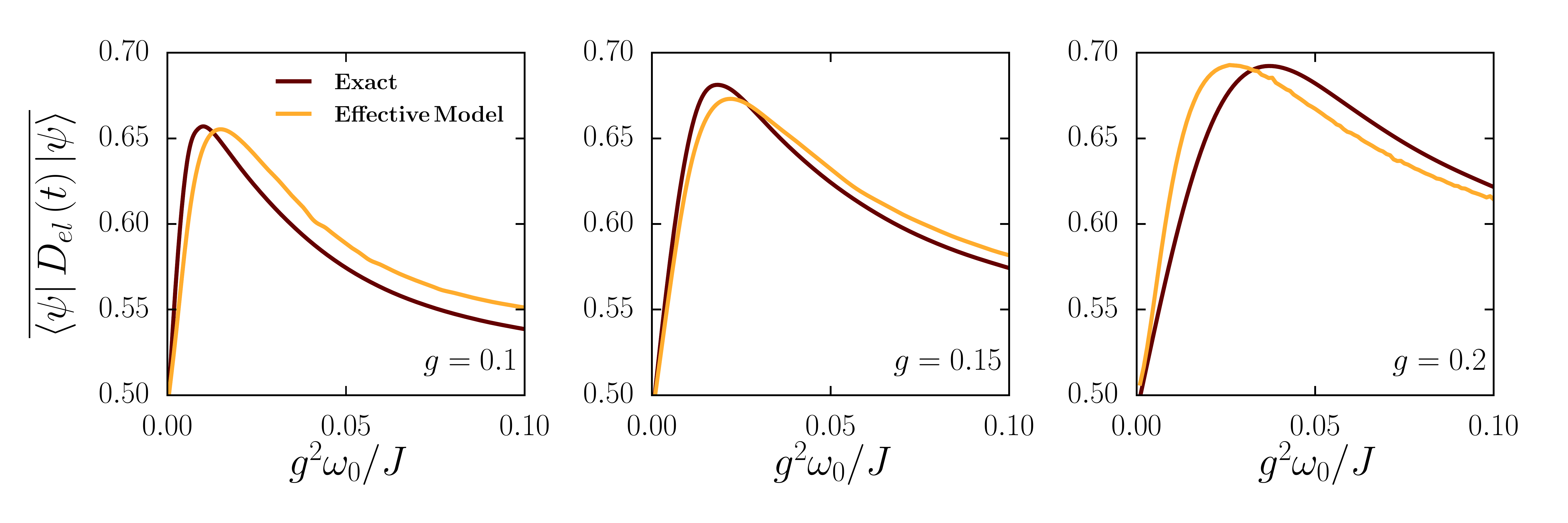}
\caption{{\bf{two-site model comparison I}}.  Comparison of the time-averaged double occupancy computed by integrating forward the exact (equation~\eqref{eq:exact_2site})  and the approximate (equation~\eqref{eq:approx_2site}) Hamiltonians from an initial state corresponding to the noninteracting two-electron ground state and a coherent state with $n_B=9$ bosons on each site using $\omega_0=1.36J$ and  values of $g$ shown. We obtained the results with 30 boson states on each site. }
\label{two_site_comp_electrons}
\end{figure}

In this subsection we investigate the electronic state. We begin with 
the  double occupancy operator
\begin{equation}
{D}_{el}=\hat{n}_{1\uparrow}\hat{n}_{1\downarrow}+\hat{n}_{2\uparrow}\hat{n}_{2\downarrow}.\label{eq:double_occ_op_def}
\end{equation}
Fig.~\ref{two_site_comp_electrons} shows the expectation value of the double occupancy averaged over the time interval from 0 to $100/J$. We compare the expectation values obtained from the exact (maroon line) and the approximate (yellow line) Hamiltonians as a function of phonon frequency at different values of coupling constant $g$.  The exact and approximate theories both produce an enhancement of the double occupancy and agree  well even at $g=0.2$. This enhancement of the double occupancy stems from (a) the induced disorder (second term of equation~\eqref{interactions}) and (b) the negative interaction induced by the phonons (third term of equation~\eqref{interactions}). The two contributions are comparable in magnitude and we thus conclude that  the quadratic electron-phonon coupling indeed produces an attractive interaction as indicated by the effective model. 

Fig.~\ref{two_site_comp_electrons} reveals a non-monotonic dependence of the double occupancy on the interaction strength $g$.  Our calculation is in essence an interaction quench and the non-monotonic dependence is a known consequence of quench dynamics. In particular, at very large values of the effective interaction it is easy to see that projecting the initial non-interacting state onto the large $|U|$ eigenstates leads to a $D\approx 0.5$, while for small enough effective interactions ($|U^\star/J^\star|< 2$, the regime of interest in this paper) the double occupancy found in the quenched problem agrees well with the one found in the corresponding ground state. 

\begin{figure}[t]
\includegraphics[width=\columnwidth]{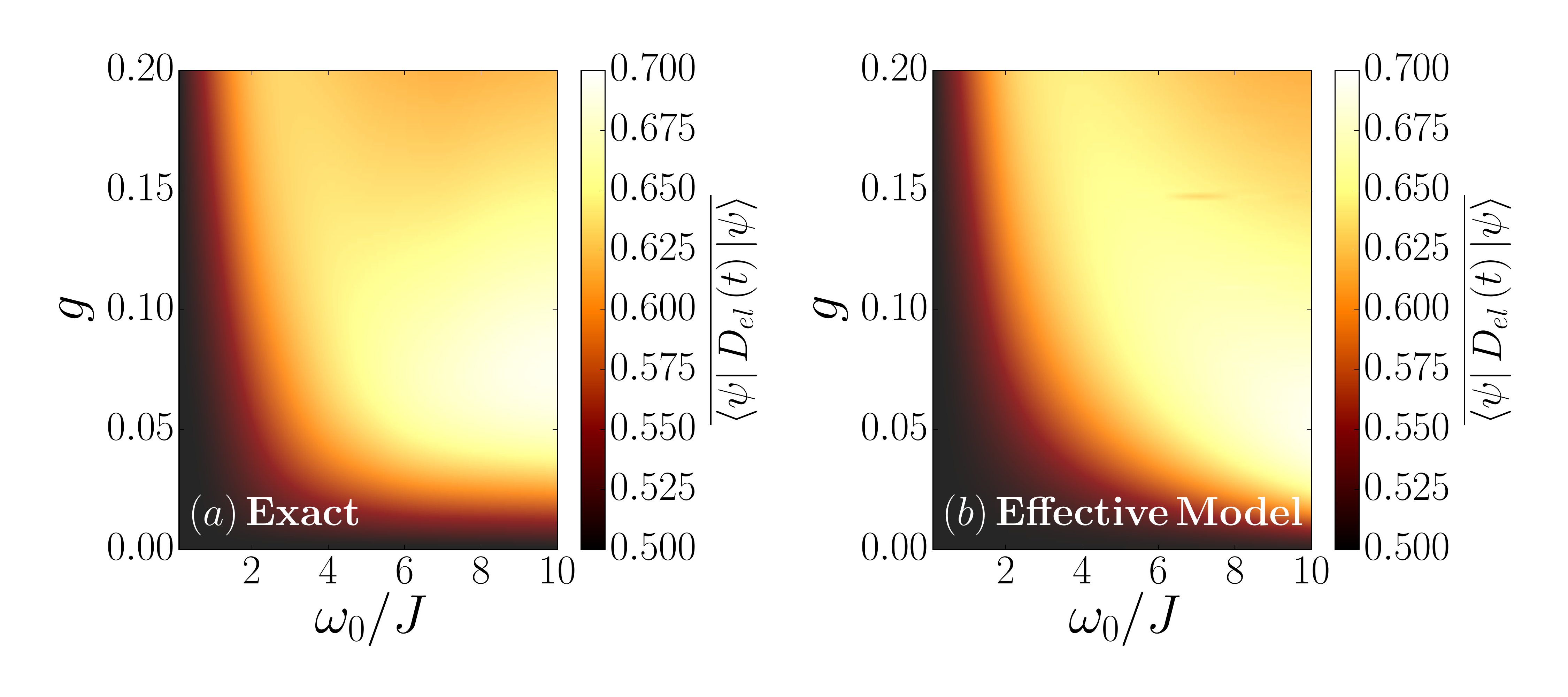}
\caption{{\bf{two-site model comparison II}}.  Comparison of the time-averaged double occupancy computed by integrating forward the exact (equation~\eqref{eq:exact_2site})  and the approximate (equation~\eqref{eq:approx_2site}) Hamiltonians from an initial state corresponding to the noninteracting two-electron ground state and a coherent state with $n_B=9$ bosons on each site as a function of  boson frequency and coupling constant. We obtained the results with 30 boson states on each site. }
\label{double_occ_wide}
\end{figure}

Finally, Fig. ~\ref{double_occ_wide} presents a comparison of the double occupancy computed in the exact and effective models, over a wide range of parameters.

\subsubsection{Spectral Function}

In this subsection we further investigate the adequacy of our approximation for the electronic dynamics by considering the electron spectral function. The computation of response functions of strongly nonequilibrium systems presents deep conceptual issues which are beyond the scope of this paper. The response functions depend in general on two time arguments: the center of time $T_{\rm c}$ and relative time $\Delta t$.  To investigate the approximations used here we consider the proxy spectral function defined by Fourier transforming with respect to $\Delta t$ and simply averaging over $T_{\rm c}$. For the electron addition part of the spectral function we have

\begin{equation}
\rho(\omega)=\sum_{m,\nu_1,\nu_2}N_{\nu_1}N_{\nu_2}\left|\left\langle m\right|c^\dagger_{1,\uparrow}\left|gs\otimes \nu_1\otimes\nu_2\right\rangle\right|^2
\delta\left(\omega-\left(E_m-E^{\nu_1,\nu_2}_0\right)\right).
\label{rhodef}
\end{equation}
Here $m$ labels the exact eigenstates of the two-site Hamiltonian and we have decomposed the initial state $\left|\Psi\right\rangle=\sum_{\nu_1,\nu_2}\sqrt{N_{\nu_1}N_{\nu_2}}\left|gs\otimes\nu_1\otimes\nu_2\right\rangle$, which was chosen as above. We also introduce the energy $E_0^{\nu_1,\nu_2}=\left\langle gs\otimes\nu_1\otimes\nu_2\right|H\left|gs\otimes\nu_1\otimes\nu_2\right\rangle$. 

In the approximate model, we additionally must keep track of the operator transformations, so we  have
\begin{equation}
\rho^{approx}(\omega)=\sum_{m,\nu_1,\nu_2}N_{\nu_1}N_{\nu_2}\left|\left\langle m\right|e^{\hat{S}}c^\dagger_{1,\uparrow}e^{-\hat{S}}\left|gs\otimes \nu_1\otimes\nu_2\right\rangle\right|^2
\delta\left(\omega-\left(E_m-E^{\nu_1,\nu_2}_0\right)\right).
\label{rhoapprox}
\end{equation}
In evaluating equation~\eqref{rhoapprox} we expanded $e^{\hat{S}}$ to $\mathcal{O}(g)$ and define $E_m,E_0$ and the eigenstates $m$ with respect to the approximate Hamiltonian.

\begin{figure}[t]
\includegraphics[width=0.85\columnwidth]{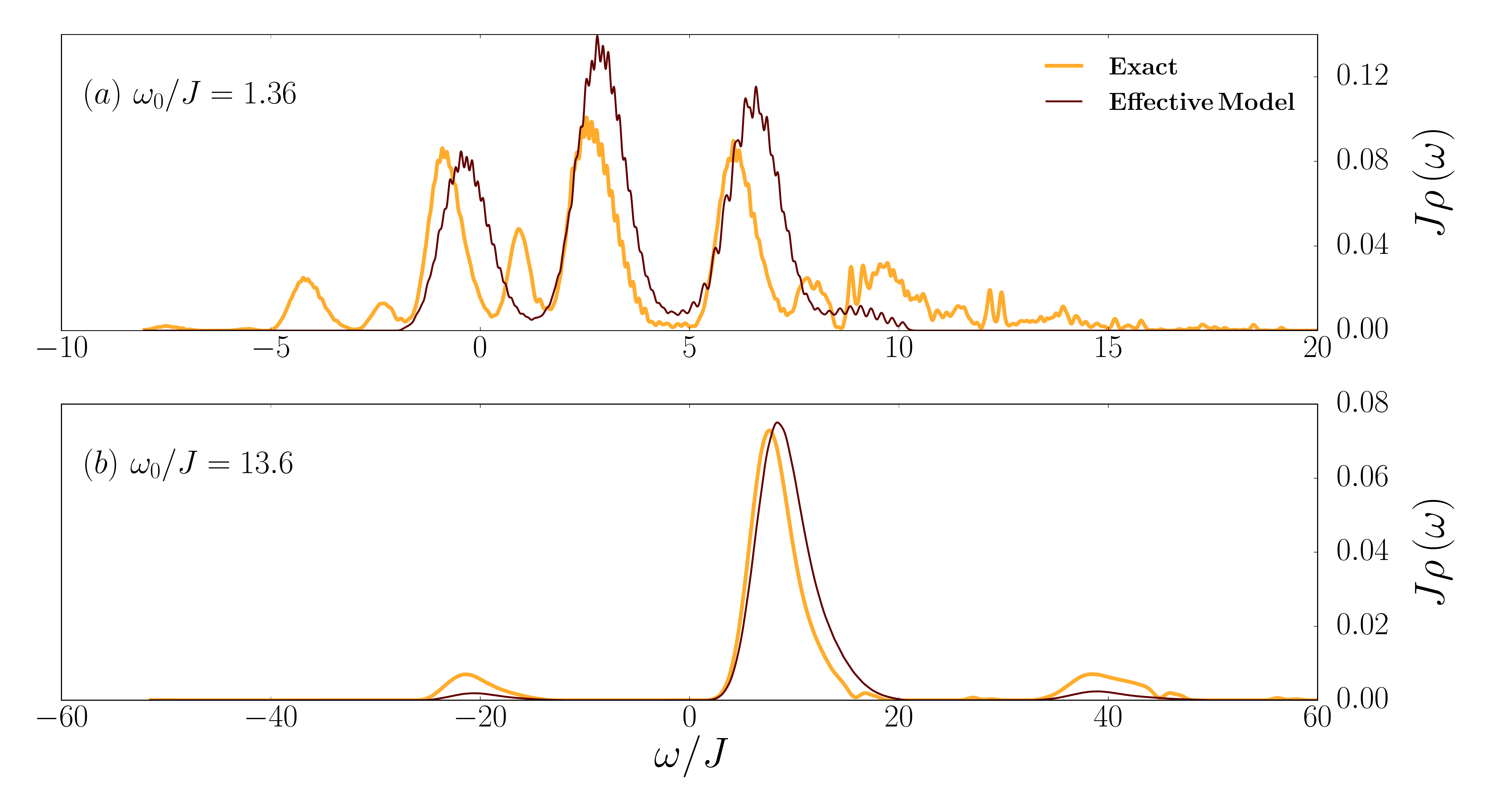}\caption{{\bf Exact and approximate spectral function} calculated from equations~\eqref{rhodef} and \eqref{rhoapprox}  for $\omega_0/J=1.36$ (a) and $\omega_0/J=13.6$ (b) and $n_{\rm B}=9$. We introduce a Gaussian broadening of the delta functions $\delta\left(x\right)\to e^{-x^2/\eta^2}/\sqrt{\eta^2\pi}$ with $\eta=0.05\omega_0$. $g$  is chosen to keep the  induced effective interaction $U^*/J= g^2\omega_0(2n_{\rm B}+1)/2 \approx 0.63$ constant. Note that properly transforming the electron creation operator $c^\dagger\rightarrow e^{\hat S}c^\dagger e^{-\hat S}$ to order $g$ is crucial here. We converged the results by including 30 boson states on each site in our numerics. }
\label{spectral}
\end{figure}

Fig.~\ref{spectral} shows a comparison of the exact and approximate spectral functions computed for two  phonon frequencies  $\omega_0/J=1.36$ (a) and $\omega_0/J=13.6$ (b). $g$  is chosen to keep the  induced effective interaction $U^*/J= g^2\omega_0(2n_{\rm B}+1)/2 \approx 0.63$ constant in both panels and we used and initial state corresponding to $n_{\rm B}=9$ in the coherent state (exact model) or decohered state (approximate). We see that evolving the electron state for fixed phonon distribution, neglecting the inelastic terms in the hopping, produces a very good approximation to the exact spectral function even for relatively low phonon frequency $\omega$.

\section{Disorder effects}

\begin{figure}[t]
\includegraphics[width=\textwidth]{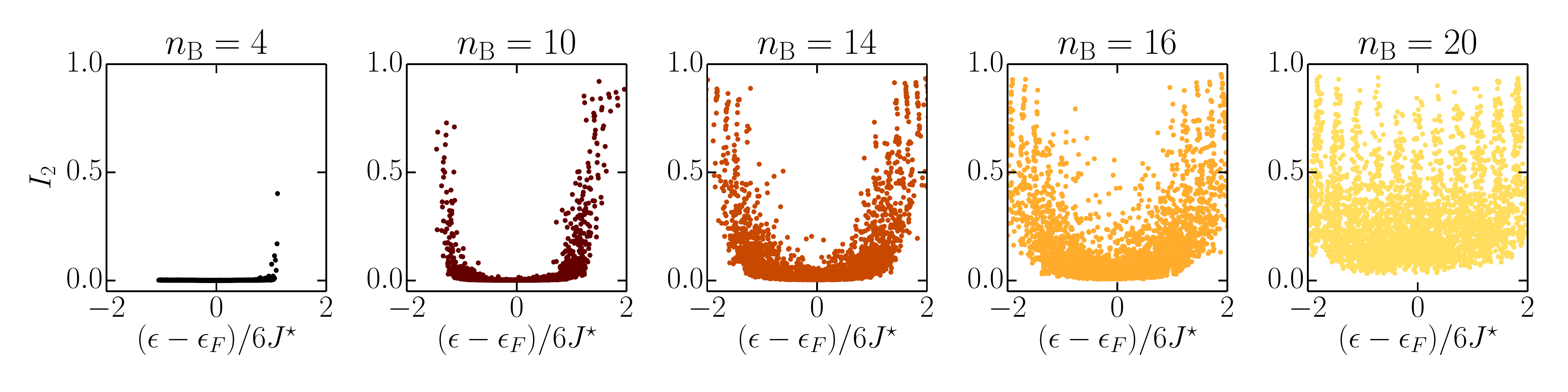}\caption{{\bf Influence of the phonon induced disorder.} The inverse partition ratio $I_{2}(\epsilon_{n})$
	is a measure of the localization of the corresponding single particle eigenstate with $I_{2}\approx0(1)$
	being extended (localized). We use a cubic lattice of $N=16^3$ lattice sites and a nearest neighbor hopping of amplitude $J^\star$ given by equation~\eqref{Jstar} as well as a disorder given by the middle term in equation~\eqref{interactions}. We choose $g=0.15$ as in the main text and vary $n_{\rm B}$. Up to $n_{\rm B}\approx 16$
	disorder has a minor effect on the localization of the states and we therefore disregard its influence in the superconductivity study of the main text (where $n_{\rm B}= 14$ per site). }
\label{fig:disorder_panels}
\end{figure}

In this section we analyse the middle term of equation~\eqref{interactions}. The analysis of the time dependence of the oscillator coordinates and electron spectral function shows that the initial coherent state evolves rapidly to a state characterized by a time-independent, incoherent, approximately Poissonian distribution of occupation numbers on each site. In this limit the middle term in equation~\eqref{interactions} corresponds to an on-site random potential, and the effect of this randomness on the properties of interest must be considered. In this paper we focus on s-wave superconductivity, which is not sensitive to moderate disorder, but is suppressed if the disorder becomes strong enough to localize the electrons. Localization cannot be assessed in the two-site model considered in previous sections. To investigate localization effects we  consider non-interacting electrons on a three dimensional cubic lattice with nearest neighbor hopping $J^\star$ with an on-site disorder modeled by the middle term of equation~\eqref{interactions} with phonon occupancies chosen from the Poisson distribution with mean occupancy $n_{\rm B}$ (the non-vanishing intersite correlations shown in Fig.~\ref{fig:coherence} means that this is a modest overestimate of the true disorder effects) and coupling strength determined by $g$. To assess localization physics we compute the participation ratio
\cite{brndiar2006universality}
\begin{equation}
I_{2}(\epsilon_{n})=\sum_{r}|\phi_{n}(r)|^{4}\label{eq:partition_ratio}
\end{equation}
for each single-particle eigenstate $n$. Extended states are characterized by a partipation ratio close to zero; localized states by a particpation ratio close to $1$.  Fig.~\ref{fig:disorder_panels} shows the participation ratio computed for $g=0.15$ (in accordance with the main text) and varying $n_{\rm B}$. we find that up to $n_{\rm B}\approx 16$ disorder has a minor effect on the localization of the states and we therefore disregard its influence in the superconductivity study of the main text (where $n_{\rm B}= 14$ per site). We note, however, that these values of $n_{\rm B}$ are not far away from the borderline where disorder should become relevant such that it might be possible to study the disorder dominated phase experimentally by increasing the fluence.

\section{Superconducting Properties}

\subsection{Variation of superconducting properties with parameters}
\begin{center}
\begin{figure}[t]
\centering
	\includegraphics[width=0.85\columnwidth]{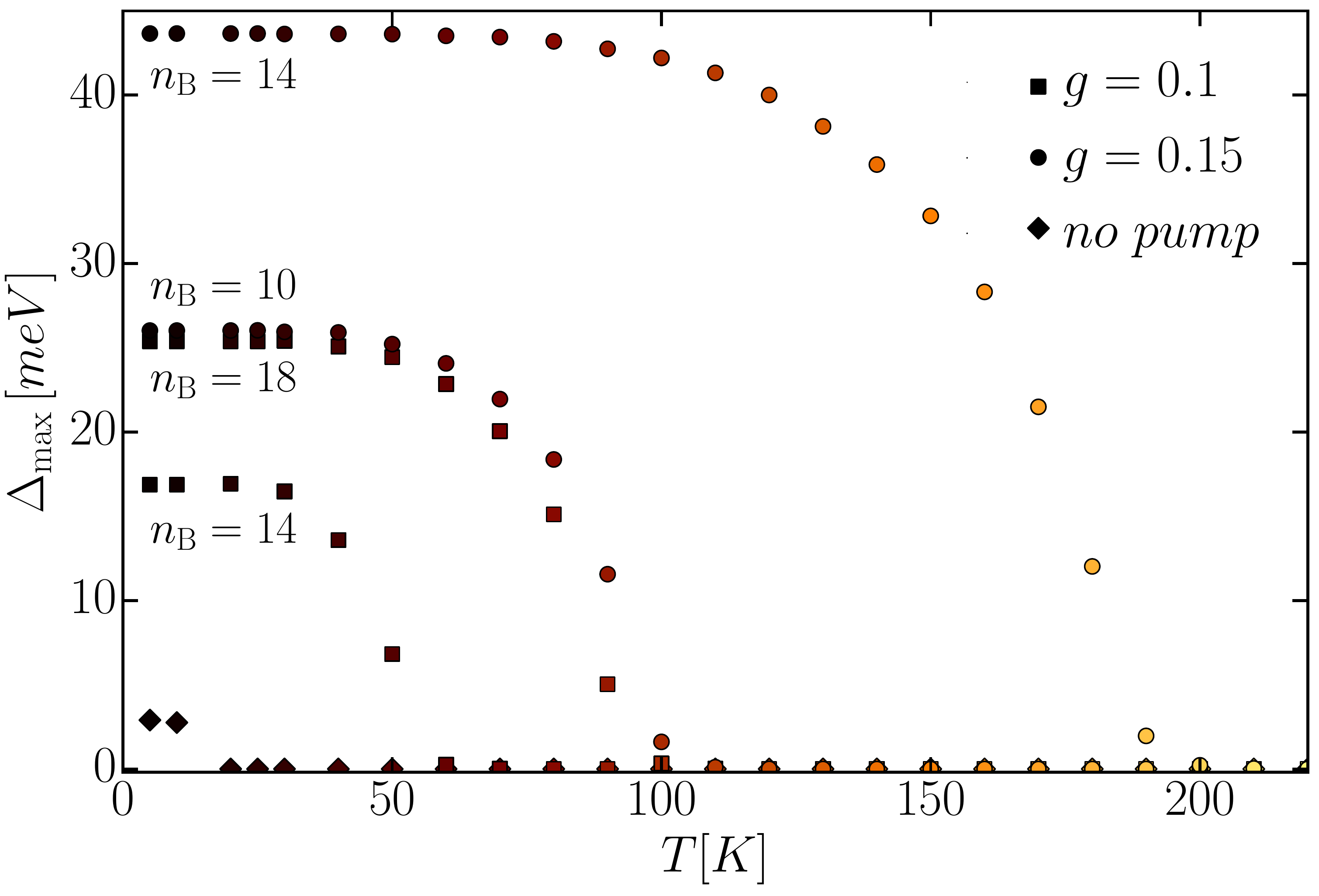}\caption{{\bf{Maximum value $\Delta_{\rm max}$ of time-dependent nonequilibrium gap function  $\Delta\left(t\right)$} }calculated as described in the main text. Different colors indicate different temperatures and circles (squares) are $g=0.15$ ($g=0.1$). For both values of $g$ we show two values of $n_{\rm B}$ as indicated in the plot. The other parameters are chosen as in Fig. 2 of the main text ($12J=0.42eV$, $\omega_0=0.17eV$ and $U$ chosen such that the equilibrium superconducting critical temperature  $T_{\rm c}^{\rm eq}=20K$). Diamonds show the equilibrium temperature dependence of the gap without the pump ($n_{\rm B}=0$) for reference.}
	\label{Fig:max_gap}
\end{figure}
\end{center}

Figure \ref{Fig:max_gap} shows the maximum value of the time dependent nonequilibrium gap function,  for different values of $g$ and mean boson excitation level $n_{\rm B}$ roughly consistent with the experiment of Mitrano {\em et al.}  We see that values of $g$ as small as $0.1$ and reasonable levels of pumping can produce significant enhancement of superconductivity. 

\subsection{Adiabatic limit}

The timescales of experimental relevance are long compared to the timescales associated with the inverse of the maximum gap or of the variation of the gap, justifying an adiabatic approximation. While the adiabatic approximation was not used to obtain the results presented for the time dependence of the gap function, it is used in the discussion of the conductivity and therefore we present here an adiabatic analysis. 

To formulate the adiabatic analysis we first introduce the time-dependent rotation $\mathbf{R}_\Delta$  that diagonalizes $\mathbf{H}(t)$ at a given instant of time $t$ (we suppress the momentum index for simplicity):
\begin{equation}
\mathbf{H}(t)=\mathbf{R}_{\Delta}^{\dagger}\tau_{3}\epsilon\left(t\right)
\mathbf{R}_{\Delta},
\label{rotatedH}
\end{equation}
with
\begin{equation}
\mathbf{R}_{\Delta}=e^{i\psi_{\Delta}(t)\left(\hat{e}_3\times\hat{n}_{\Delta}\right)\cdot\vec{\tau}}.\label{Rdeltadef}
\end{equation}

In the above, $\tau_{i}$ are Pauli matrices, $\hat{n}_{\Delta}$ is the unit vector in the superconducting ($1-2$)  subspace that is parallel to ${\rm Re}[\Delta(t)]\tau_{1}+{\rm Im}[\Delta(t)]\tau_{2}$ and $2\psi_{\Delta}$ is the rotation angle in Nambu space which rotates the instantaneous time-dependent Hamiltonian to a diagonal form with eigenvalues
\begin{equation}
\epsilon(t)=\sqrt{\varepsilon_{k}^{2}(t)+\Delta^{2}(t)}.\label{Edef}
\end{equation}
Writing 
\begin{equation}
\mathcal{U}\left(t\right)=\mathbf{R}_{\Delta}\left(t\right)\mathbf{R}^{\dagger}\left(t\right),
\end{equation}
and inserting the result into the equations of motion (see methods section of main text) we obtain
\begin{equation}
-i\mathcal{U}^{\dagger}\left(t\right)\dot{\mathcal{U}}\left(t\right)-i\partial_{t}\log\mathbf{D}\left(t\right)=\epsilon\left(t\right)\mathcal{U}^{\dagger}\left(t\right)\tau_{3}\mathcal{U}\left(t\right)-\dot{\psi}_{\Delta}\left(t\right)\mathcal{U}^{\dagger}\left(t\right)\tau_{2}\mathcal{U}\left(t\right).\label{gap_eq_adia}
\end{equation}
In the adiabtic limit we expect $\mathcal{U}\left(t\right)-\mathbf{1}\ll\mathbf{1}$
as are all time derivatives of $\mathbf{D}.$ One can then write
\begin{equation}
\mathcal{U}\left(t\right)=e^{i\vec{v}\left(t\right)\cdot\vec{\tau}}\approx\mathbf{1}+i\vec{v}\left(t\right)\cdot\vec{\tau},
\end{equation}
which gives $\mathcal{U}^{\dagger}\left(t\right)\dot{\mathcal{U}}\left(t\right)=i\dot{\vec{v}}\cdot\vec{\tau}$
and $\mathcal{U}^{\dagger}\left(t\right)\tau_{3}\mathcal{U}\left(t\right)=\tau_{3}-2\left(\hat{z}\times\vec{v}\right)\cdot\vec{\tau}-\dot{\psi}_{\Delta}\tau_{2}.$
To this order we have
\begin{figure}[t]
\includegraphics[width=0.85\columnwidth]{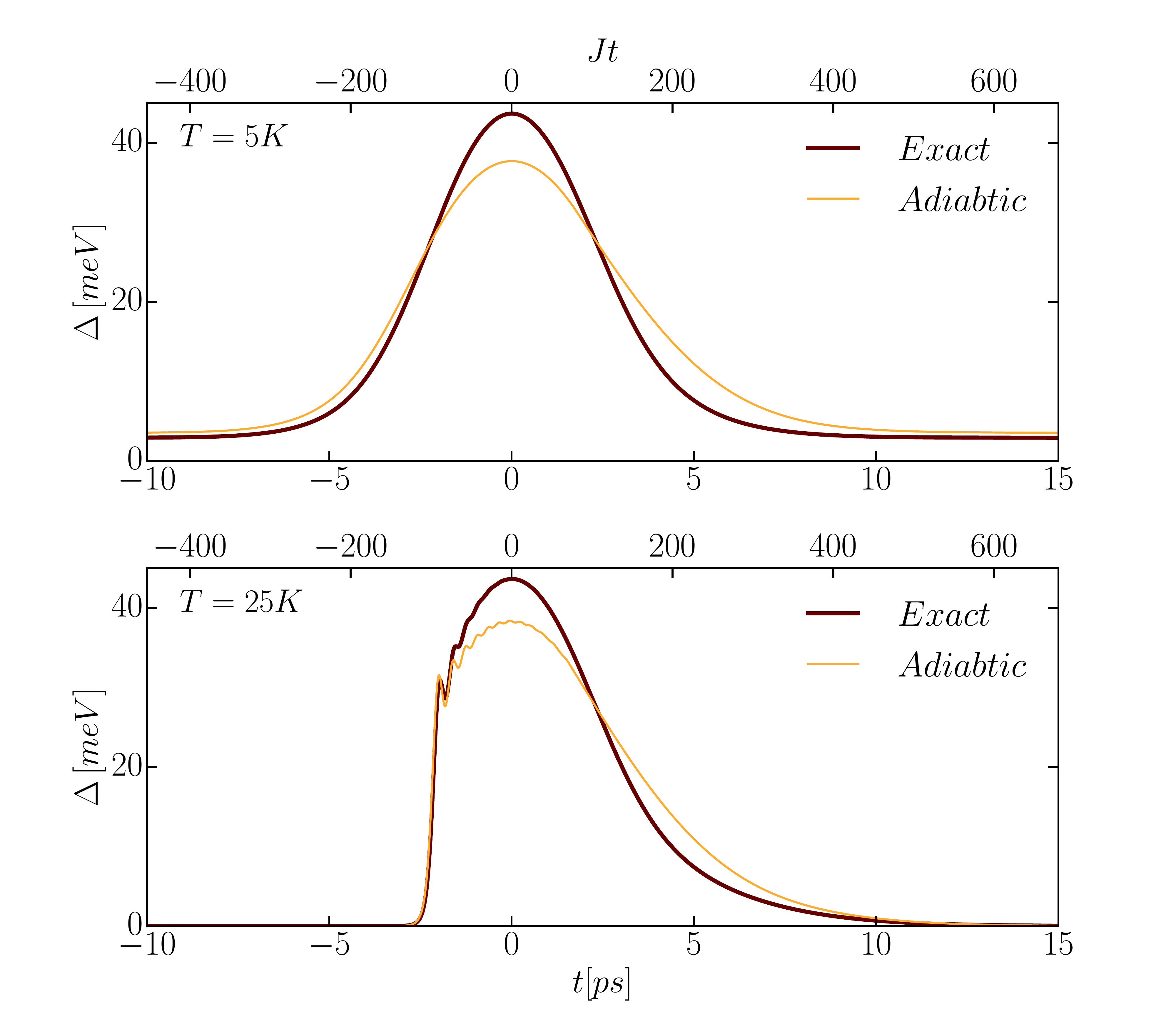}\caption{{\bf{Exact vs. adiabtic time dependent gap. }}Comparison between the exact solution of the gap equation given by
equation~(11) of the main text and the one resulting from the adiabtic
limit in equation~\eqref{gap_eq_adia} for both an initially
superconducting state $\left(T=5K\right)$ and a normal state
$\left(T=25K\right).$ The calculation is done with the same parameters as Fig.~2 of the main text. }
\label{Fig:adiabtic}
\end{figure}
\begin{equation}
\mathbf{D}\left(t\right)=e^{-i\tau_{3}\int_{0}^{t}\epsilon\left(s\right)\,ds},
\label{dadiabatic}
\end{equation}
while $\vec{v}$ lies in the $1-2$ plane and rotates as
\begin{equation}
\dot{v}_{1}\left(t\right)-2\epsilon\left(t\right)v_{2}\left(t\right)=0,
\end{equation}
\begin{equation}
\dot{v}_{2}+2\epsilon\left(t\right)v_{1}\left(t\right)=-\dot{\psi}_{\Delta}.
\end{equation}
Once we obtain $\vec{v}\left(t\right)$ we re-exponentiate to preserve
the unitary properties of $\mathcal{U}\left(t\right).$ In Fig. \ref{Fig:adiabtic}
we show the comparison between the exact time-dependent gap and the
one computed via the adiabtic approximation. These results indicate
that the system is indeed in the adiabtic limit for parameters relevant to the experiment. 

The optical conductivity calculations use this result with the additional approximation that the integral of $\varepsilon(s)$ in equation~\eqref{dadiabatic} is replaced by the value $t\varepsilon(t/2)$.

\newpage
%

\end{document}